\documentclass[twocolumn,showpacs,floatfix,amsmath,amssymb,prb]{revtex4}
\usepackage[dvips]{graphicx}
\usepackage{latexsym}
\usepackage{graphicx}
\usepackage{times,psfrag,subfigure}
\usepackage{amsmath}
\usepackage{dcolumn}% Align table columns on decimal point
\newcommand{\be}{\begin{equation}}
\newcommand{\ee}{\end{equation}}
\newcommand{\ba}{\begin{eqnarray}}
\newcommand{\ea}{\end{eqnarray}}
\newcommand{\baa}{\begin{eqnarray*}}
\newcommand{\eaa}{\end{eqnarray*}}

\begin{document}

\title{Spin pumping and magnetization dynamics in ferromagnet-Luttinger liquid junctions }
\author{Cristina Bena}
\email{cristina@physics.ucsb.edu}
\author{Leon Balents}
\email{balents@physics.ucsb.edu}
\affiliation{Department of Physics, University of California at Santa Barbara,
Santa Barbara, CA-93106}
\date{\today}

\begin{abstract}
We study  spin transport between a ferromagnet (FM) with 
time-dependent magnetization and a conducting carbon nanotube or
quantum wire, modeled as a Luttinger liquid (LL). 
The precession of the magnetization vector of the ferromagnet due for instance to
an outside applied magnetic field causes spin  pumping 
into an adjacent conductor. Conversely, the spin injection causes increased magnetization
damping in the ferromagnet.  We find that, if the conductor adjacent to the 
ferromagnet is a Luttinger liquid, 
spin pumping/damping is suppressed by interactions, and the suppression has clear
Luttinger liquid power law temperature dependence.
We apply our result to a few particular setups.
First we study  the effective Landau-Lifshitz-Gilbert (LLG)
coupled equations for the magnetization vectors of the 
two ferromagnets in a FM-LL-FM junction. Also,
%along the lines of Refs.~\onlinecite{Brataas1,Brataas2} 
we compute the Gilbert 
damping for a FM-LL and a FM-LL-metal junction.
\end{abstract}
\maketitle

\section{Introduction}
In the past years there has been a lot of interest in the 
field of spintronics,
due especially to the many possible applications of spin-polarized transport.
However, practical realization of devices in which spin can be well
controlled and which allow for large stable spin currents  is difficult.
For instance, spin injection from magnetic to normal materials is reduced 
due to a large resistivity mismatch \cite{inj}, while the 
spin-polarized currents obtained through optical pumping 
\cite{optical} are small.
Following the ideas of adiabatic charge pumping in mesoscopic systems 
\cite{pump}, Y. Tserkovnyak {\it et. al.} proposed  a new method of obtaining
spin polarized currents through injection
of spin from a ferromagnet with time dependent magnetization into 
a normal metal\cite{Brataas1,Brataas2,Brataas3}. 
The advantage of this method lies in the possibility of  
direct significant spin current injection. 
The authors also showed that 
spin pumping has important experimentally observable 
effects on the dynamics of the ferromagnet.
In general, the dynamics of a magnetization vector in a ferromagnet is 
described by the LLG equation \cite{Gilbert}. This includes a term characterizing the precession
of the magnetization vector 
in an external magnetic field, as well as damping terms which describe
the reduction of the magnetization due to spin-flip processes, etc. The coefficient
of the damping term is called ``Gilbert damping'', and can be determined
experimentally. The  spin loss in the ferromagnet 
due to spin pumping, generates a renormalization of 
the coefficients in the LLG equation.  Of most experimental interest is 
the renormalization of the Gilbert damping  coefficient.
The magnitude of this effect is 
estimated in Refs.~\onlinecite{Brataas1} and \onlinecite{Brataas2} for spin pumping from
a FM into a normal metal.

In this work we focus on spin pumping from a ferromagnet into
a Luttinger liquid. For this setup other theoretically interesting 
questions also arise, like what is the effect of the electron-electron
interactions in the LL on the injection of
spin, and how the interactions  will affect the renormalization 
of parameters such as the Gilbert damping. We find that the spin injection
and the
renormalization of the Gilbert damping are suppressed due to interactions,
and that the suppression has a power law dependence on temperature.

The study of this setup is also important due to its possible applications; 
given the excellent transport properties of some LL's,
especially of carbon nanotubes, this setup
would open the perspective of transporting the injected spin over
large distances without important losses. The study of spin pumping into a Luttinger liquid is also
of more general interest as a prototype for the effect of Coulomb interactions (i.e. charging 
effects ) on spin injection.

In section II we
compute the spin current pumped into a LL from a ferromagnet
with a precessing magnetization. In section
III we apply this result to study a FM--LL--FM junction assuming that the transport in the LL is 
ballistic. In section IV
we compute the renormalization of the Gilbert damping parameter
due to spin pumping from a FM into 
a finite size  Luttinger liquid with diffusive transport. 
In section 
V we repeat this analysis for a FM-- LL-- metal junction
assuming that the transport in LL is ballistic. 
We discuss the results and conclude in
section VI.

\section{Spin pumping between a ferromagnet with time dependent magnetization 
and a Luttinger liquid}

We focus first on the spin current pumped from a FM 
into a LL as a result of the precession in the FM's 
magnetization vector. 
We assume almost perfect backscattering,
and a small amount 
of tunneling between the LL and the FM. The corresponding boundary conditions
are $\Psi_{R \alpha}(x=0)=\Psi_{L \alpha}(x=0) =\Psi_{\alpha}\propto
e^{i \phi_{\alpha}(x=0)}$, and similarly 
$F_{R \alpha}(x=0)=F_{L \alpha}(x=0) 
=F_{\alpha}
\propto
e^{i \phi_{\alpha}^m(x=0)}$. Here the operators $\Psi/F_{L/R \alpha}$ 
correspond to the
annihilation of a left/right moving electron with spin 
$\alpha=\uparrow/\downarrow$ in the LL/FM, while $\phi_{\alpha}$ 
and $\phi_{\alpha}^m$ respectively are the corresponding 
bosonized operators in the LL/FM sectors. 
The  effective actions in bosonized variables for the
 LL and the FM are given by \cite{KF}
\ba 
&&S_{LL}=\sum_{\omega_n}\sum_{a=\rho/\sigma}
\frac{g_a \beta |\omega_n|}{\pi} |\phi_a(\omega_n)|^2,
\nonumber \\&&
S_{FM}=\sum_{\omega_n}\sum_{a=\rho/\sigma}
\frac{\beta |\omega_n|}{\pi} |\phi^a_m(\omega_n)|^2,
\ea
where $\beta$ is inverse temperature, $\phi_{\rho/\sigma}=(\phi_{\uparrow}\pm
\phi_{\downarrow})/2$, $g_{\sigma}=1$ and $g_{\rho}=g$ is the interaction 
parameter in the LL.

In terms of the fermionic operators the tunneling Hamiltonian can be 
described by

\ba
H_{tun}&=&\sum_{\alpha=\uparrow/\downarrow}
(u_1 F_{\alpha}^{\dagger} \Psi_{\alpha}+h.c.)
\nonumber \\&&
+ \vec{m}(t) \! \cdot \!
\sum_{\alpha/\beta=\uparrow/\downarrow}(u_2 F_{\alpha}^{\dagger}  
\frac{\vec{\sigma}_{\alpha \beta}}{2}
\Psi_{\beta}+h.c.),
\ea
where $\vec{m}(t)$ is the time dependent magnetization in the ferromagnet in units of
$\hbar$.

The spin current operator corresponding to tunneling
into the Luttinger liquid is
\be
\vec{I}=\frac{d}{d t}\bigg[\int dx \vec{M}(x)\bigg]_{tun}=
\int \frac{dx}{\hbar}(-i [\vec{M}(x),H_{tun}]),
\ee
where $\vec{M}(x)=\hbar \sum_{\alpha/\beta=\uparrow/\downarrow}
\Psi_{\alpha}^{\dagger} (x) 
\frac{\vec{\sigma}_{\alpha \beta}}{2}
\Psi_{\beta}(x)$ is the spin density in the LL at position $x$. 
Consequently this can be written as
\ba
\vec{I}&=&i(\frac{u_1}{2} F^{\dagger}_{\alpha} \vec{\sigma}_{\alpha \beta}
\Psi_{\beta}+\frac{u_2}{4} \vec{m} F_{\alpha}^{\dagger} \Psi_{\alpha}-h.c)
\nonumber \\&&+(\frac{u_2}{4} \vec{m} \times \vec{\sigma}_{\alpha \beta}
F^{\dagger}_{\alpha} \Psi_{\beta}+h.c.),
\ea 
where sums over repeated spin indices $\alpha$ and $\beta$ are implied.

We compute the spin current 
using time dependent perturbation theory $\langle I(t) \rangle=(-i/\hbar)
\int dt' \Theta(t-t') \langle [I(t),
H_{tun}(t')]\rangle$:

\begin{widetext}
\ba
 \langle\vec{I}(t)\rangle &=&-\int \frac{dt'}{\hbar}\{ 
\vec{m}(t) {\rm Im}[u_1^{*} u_2 C(t-t')]+\vec{m}(t') {\rm Im}[u_2
^{*} u_1 C(t-t')]\}
-\frac{1}{2} |u_2|^2 \int \frac{dt'}{\hbar}
 \vec{m}(t')\times\vec{m}(t) {\rm Re}[C(t-t')]
\nonumber \\
&=&\frac{\vec{m}(t)}{\hbar} {\rm Im}(u_2^{*} u_1) {\rm Re} [C(0)]
-\int \frac{d \omega}{h} C(\omega)e^{-i \omega t}[{\rm Im}(u_2^{*} u_1)
\vec{m}(\omega)+\frac{1}{2}|u_2|^2\vec{m}(\omega) \times \vec{m}(t)]
\ea
\end{widetext}
where $C(t-t')=-i \Theta(t-t') \langle[F^{\dagger}_{\alpha} 
(t)\Psi_{\alpha}(t),
\Psi^{\dagger}_{\alpha}(t') F_{\alpha}(t')]\rangle$ is the 
retarded Green's function, $C(\omega)=\int (d \omega/2 \pi) C(t) e^{i \omega t}$ is the Fourier
transform of $C(t)$,
and $C(0)=C(\omega=0)$.

%{\bf this formula does not yield Eq.(9) but it produces a minus %sign in front of the first term 
%$A_1$}.

We assume that the terms proportional to ${\rm Re}[C(\omega)]$
 drop from $I(t)$. The function ${\rm Re}[C(\omega)]$ is even with respect 
to frequency. Since
we are interested in frequencies smaller than other parameters 
in the system (temperature, etc.) we can Taylor 
expand ${\rm Re}[C(\omega)]$ and keep only 
the lowest order contributions. However the zeroth order expansion terms cancel, 
while the next non-zero higher 
order terms are proportional to $\omega^2$, and can be dropped from the current.

Also we can write 
\cite{spin1,spin2,spin3},
\ba
{\rm Im}[C(\omega)]&\propto& 
\frac{1}{k_B T } \Big(\frac{k_B T}{\epsilon0}\Big)^{\delta+2}
\sinh \big(\frac{\hbar 
\omega}{2 k_B T}\big)\times
\nonumber \\&& \times \Big|\Gamma\Big(1+\frac{\delta}{2}+
i\frac{\hbar \omega}{2 \pi k_B T}\Big)\Big|^2,
\ea
where $\delta=(1/g+1)/2-1$
for the case of a LL of interaction parameters $g_{\rho}=g$, $g_{\sigma}=1$ ,
and  $\delta=(1/g+3)/4-1$ for the case of a nanotube for which a typical $g \sim 0.25$.
Also, $\epsilon_0$ is a high energy cutoff, and for a nanotube  $\epsilon_0
\sim 1{\rm eV}$ is the energy level spacing. 

For $\hbar \omega/k_B T \ll 1$

\be
{\rm Im}[C(\omega)]\propto 
\hbar \omega 
\Big(\frac{k_B T}{\epsilon_0}\Big)^{\delta} \big|\Gamma \big(1+\frac{\delta}{2}
\big)\big|^2,
\ee
 and for $\hbar\omega/k_B T \gg 1$

\be
{\rm Im}[C(\omega)]\propto \hbar \omega
\Big(\frac{\hbar \omega}{2 \pi \epsilon_0}\Big)^{\delta}.
\ee
For  temperatures of the order of $100K$ and reasonable frequencies 
achievable experimentally
we can safely assume that we are in the limit 
of $\hbar\omega/k_B T \ll 1$, so that we can use the first expression above. 
Consequently
the injected spin current has the same form as for 
non-interacting fermions \cite{Brataas1,Brataas2}
but with temperature dependent coefficients.
\ba
I(t) &=& - A_1
\Big(\vec{m}\times \frac{d \vec{m}}{dt}\Big)+
A_2 \frac{d \vec{m}}{dt}, \text{   with} 
\nonumber \\
A_1&=&\frac{A_r}{2}\Big(\frac{T}{\epsilon_0}\Big)^{\delta},
\nonumber \\
A_2&=&\frac{A_i}{2}\Big(\frac{T}{\epsilon_0}\Big)^{\delta},
\label{i}
\ea
where $A_r \propto  |u_2|^2$ and 
$A_i \propto  {\rm Im}[u_1^{*} u_2]$.

This is the main result of this paper. In the following sections we will apply this 
result to a few particular setups.

\section{Ferromagnet -- Luttinger liquid -- Ferromagnet junction}

We first consider a ferromagnet -- Luttinger liquid -- ferromagnet junction. The 
magnetizations of the two ferromagnets are time dependent, and
spin currents are injected into the wire from both ends.

\begin{figure}[h]
\begin{center}
\includegraphics[width=2.5in]{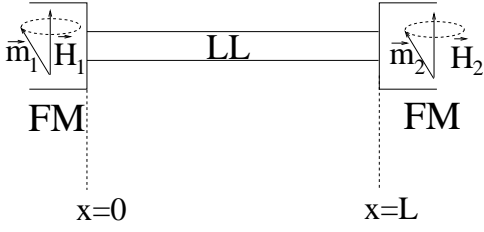}
\end{center}
\vspace{0.15in}
\caption{Ferromagnet-Luttinger liquid-ferromagnet junction}
\label{fig3}
\end{figure}

We assume that the two
contacts are identical, 
and that the transport through the wire is ballistic, so that   
the spin current $\vec{I}$ and the spin chemical potential $\vec{\mu}_s$ are uniform 
throughout the wire. 

The coupled Landau-Lifshitz-Gilbert equations\cite{Brataas1,Gilbert} for the 
magnetization vectors in the two ferromagnets can be written as:

\be
\frac{d \vec{m}_i}{dt}=-{\gamma} \vec{m}_i\times \vec{H_i}+
\alpha_0 \vec{m}_i\times
\frac{d \vec{m}_i}{dt}\mp\frac{\gamma}{M_s} \vec{I}
\label{mi}
\ee
where ${\gamma}=g \mu_B/\hbar$, $g$ is the gyromagnetic ratio,
$\mu_B$ is the Bohr magneton, 
$\alpha_0$ is the dimensionless bulk Gilbert 
damping constant and $M_s$ is the ferromagnetic saturation magnetization. 
The index $i=1,2$ characterizes the two ferromagnets. Our convention is
that  spin current  flows from FM $1$ to FM $2$. Thus the $-$ sign in front of the
term proportional to the spin current
in Eq. (\ref{mi}) corresponds to spin injection from FM $1$ 
into the wire, while the $+$ sign
corresponds to spin current flowing from the wire
into FM $2$.

The total spin current flowing through the wire
$\vec{I}=\vec{I}^1_0-\vec{I}_b^1=-\vec{I}_0^2+\vec{I}_b^2$ 
has two contributions. The terms $\vec{I}_0^i$ denote the spin currents pumped from
the two ferromagnets solely as a result of the time dependence in the magnetization vectors. 
The general
form for these currents has been derived in the previous section. As noted, their 
magnitudes depend only on the nature of the junctions, and are independent of setup 
details. The second contribution, the backscattered currents $\vec{I}_b^i$,
are the result of a spin flow back to the ferromagnets due to spin accumulation
in the Luttinger liquid. Since how much spin accumulates on the wire depends on 
various setup parameters, the backscattered currents also depend on 
various parameters such as the length of the wire, the spin flip time in the wire, etc.
In general, as has been derived in Refs.~\onlinecite{spin1} and \onlinecite{spin3}

\be
\vec{I}_b^i=
\frac{\theta}{4 \pi} \vec{\mu}_s \times \vec{m_i}-\frac{\vec{\mu}_s}{\mu_s}
I_{\delta}(\mu_s,T).
\label{ib}
\ee
Here $\theta$ is a dimensionless parameters which 
measures the exchange coupling  between the FM and the LL \cite{spin1,spin3}, and
\ba
&&I_{\delta}(\mu_s,T)\propto (4 |u_1|^2+|u_2|^2)k_B T 
(k_B T/\epsilon_0)^{\delta} \times
\nonumber \\&&
\times \sinh \big(\frac{\mu_s}{4 k_B T}\big)
\Big|\Gamma\Big(1+\frac{\delta}{2}+i\frac{\mu_s}{4 \pi k_B T}\Big)\Big|^2.
\ea
The spin chemical potential in the wire is related to the magnetization by
$\vec{\mu}_s(x)=\vec{M}(x)/\chi$, where $\chi=1/(2 \pi v)$ is the spin susceptibility
in the LL. 
For the rest of the paper we assume that the condition  $\mu_s \ll k_B T$ 
is satisfied, and we  take  $I_{\delta}(\mu_s,T)/\mu_s$
to be a coefficient independent of $\mu_s$. We denote it by
${\cal T}\propto (4 |u_1|^2+|u_2|^2|) 
(k_B T/\epsilon_0)^{\delta}$.

From the above equations, the spin current is found to be
\begin{widetext}
\ba
\vec{I}=&&\frac{\vec{I}_0^1-\vec{I}_0^2}{2}
%\nonumber \\&&
+\frac{ \theta^2}{4(32 \pi^2 {\cal T}^2+\theta^2+\theta^2 
\vec{m}_1 \cdot \vec{m}_2)} \Big[\frac{8 \pi {\cal T}}
{\theta}(\vec{I}_0^1+\vec{I}_0^2)
\times (\vec{m}_1-\vec{m}_2)
%\nonumber \\&&
+(\vec{m}_1+\vec{m}_2)
(\vec{I}_0^2\cdot\vec{m}_1- \vec{I}_0^1
\vec{m}_2) 
\nonumber \\&&
+\frac{\theta}{8 \pi \cal T} (\vec{m}_2 \times \vec{m}_1) 
(\vec{I}_0^2\cdot\vec{m}_1+ \vec{I}_0^1\vec{m}_2) \Big],
\label{is}
\ea
\end{widetext}
where we assumed $m_1^2=1$
and $m_2^2=1$.

A tractable limit of the problem is the  case of negligible exchange terms,
$\theta=0$. A detailed analysis of this situation is given in Appendix A. 
Assuming that the two ferromagnets are subject to identical magnetic fields
$\vec{H}_1=\vec{H}_2$, and linearizing the LLG equations we obtain
\begin{widetext}
\ba
m_x^{1/2}&=&\frac{m_1+m_2}{2}\cos\frac{\gamma H t}{1+\alpha_0^2}
\exp\Big(-\frac{\alpha_0 \gamma
H t}{1+\alpha_0^2}\Big)
\nonumber \\&&
\pm \frac{m_1-m_2}{2}\cos\frac{(1+2 \alpha_2)\gamma H t}{(\alpha_0+
2\alpha_1)^2+(1+2 \alpha_2)^2}
\exp\Big[-\frac{(\alpha_0 +2 \alpha_1)\gamma H t}{(\alpha_0+
2\alpha_1)^2+(1+2 \alpha_2)^2}\Big],
\nonumber \\
m_y^{1/2}&=&\frac{m_1+m_2}{2}\sin\frac{\gamma H t}{1+\alpha_0^2}
\exp\Big(-\frac{\alpha_0 \gamma
H t}{1+\alpha_0^2}\Big)
\nonumber \\&&
\pm \frac{m_1-m_2}{2}\sin\frac{(1+2 \alpha_2)\gamma H t}{(\alpha_0+
2\alpha_1)^2+(1+2 \alpha_2)^2}
\exp\Big[-\frac{(\alpha_0 +2 \alpha_1)\gamma H t}{(\alpha_0+
2\alpha_1)^2+(1+2 \alpha_2)^2}\Big],
\ea
\end{widetext}
where $\alpha_{1/2}=\gamma A_{1/2}/2 M_s$, and
we assumed the initial conditions, $m_{1/2}^x=m_{1/2}$ and $m_{1/2}^y=0$.
We note that the solutions contain two terms, each of them being
a superposition of precession and exponential decay. The first 
term $(m_1+m_2)$ is not affected in any way by the Luttinger liquid physics, while
in the second term ($m_1-m_2$) the time constants for precession and decay are
modified due to the LL interactions. In particular, power law dependencies on temperature
are added to these coefficients through the $\alpha_{½}$ terms.

\section{Renormalization of the Gilbert damping constant due to the 
spin flow between a ferromagnet and a Luttinger liquid}

As before, the LLG equation for the magnetization in the FM can be written as

\be
\frac{d \vec{m}}{dt}=-{\gamma} \vec{m}\times \vec{H}+
\alpha_0 \vec{m}\times
\frac{d \vec{m}}{dt}-\frac{\gamma}{M_s} \vec{I},
\label{llg}
\ee
where, similar to the previous
section, the spin current $\vec{I}=\vec{I}_0-\vec{I}_b$ is the total 
current flowing out of the ferromagnet.
The two contributions $\vec{I}_0$ and $\vec{I}_b$ correspond to the
pumped and backscattered spin currents, and are described by Eqs.
(\ref{i}), and (\ref{ib}) with $\vec{m}_i=\vec{m}$, respectively.
Equation (\ref{llg}) can be rewritten as

\be
\frac{d \vec{m}}{dt}=-\gamma' \vec{m}\times \vec{H}+\alpha \vec{m}\times
\frac{d \vec{m}}{dt},
\ee
where $\gamma'$ and ${\alpha}$ are renormalized constants.
Our goal is to compute the total current $\vec{I}$ and consequently 
determine the coefficients $\gamma'$ and ${\alpha}$ for various setups.

In this section we compute these renormalized constants for a
 ferromagnet in contact with a LL of finite size $L$, and 
characterized by
a spin flip time $\tau_{sf}$.  We assume that the transport in the LL at long 
wavelengths is diffusive.

\begin{figure}[h]
\begin{center}
\includegraphics[width=2.5in]{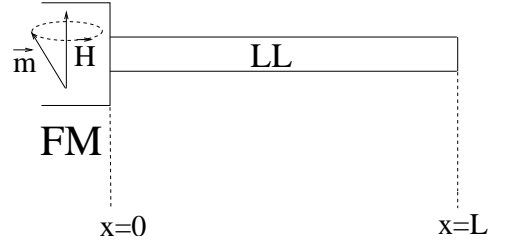}
\end{center}
\vspace{0.15in}
\caption{Ferromagnet-Luttinger liquid junction}
\label{fig1}
\end{figure}

A few comments are in order: if $\tau_{sf}$ is very large, i.e. 
if the spin flip is
negligible, after a long enough time the spin accumulation in the 
LL would be large, such that the current injected from the FM would equal 
the backscattered 
current, and the total current will be zero. If $\tau_{sf}$ is very small, 
any spin accumulation in the LL is relaxed instantaneously, the backscattered 
current is negligible and the total current is equal to the injected current.

For a real system neither situation may be correct, and we take
the spin flip time $\tau_{sf}$ to be finite. Along the lines 
of  Ref.~\onlinecite{Brataas1} the diffusion 
equation for the spin in the LL is
\be
i \omega \vec{\mu}_s= D \partial_x^2 \vec{\mu}_s-\tau_{sf}^{-1} \vec{\mu}_s,
\ee
with the boundary conditions
$\partial_x \vec{\mu}_s(x)=- (2 \pi v_F/D)  \vec{I}$,
at $x=0$,
and of  vanishing spin current, $\partial_x 
\vec{\mu}_s(x)=0$, at $x=L$. Here $D$ is the diffusion coefficient.

The coupled equations for $\vec{\mu}_s(x)$ and $\vec{I}$ can be solved (for 
details see Appendix B). 
The renormalized LLG equation coefficients are:
\ba
\gamma'&=&\frac{\gamma}{1+\frac{\gamma}{M_s}(B_1 A_2-B_2 A_1)},
\nonumber \\ 
\alpha&=&\frac{\alpha_0+\frac{\gamma}{M_s}(B_1 A_1+ A_2  B_2)}{1+
\frac{\gamma}{M_s}(B_1 A_2-B_2 A_1)}.
\ea
where $B_1=
(1+\xi {\cal T}) /[(1+\xi {\cal T})^2+(\xi \theta)^2/(16 \pi^2)]$,
$B_2=(\xi \theta)/\{4 \pi [(1+\xi {\cal T})^2+(\xi \theta)^2/(16 \pi^2)]\}$, and
$\xi=(2 \pi v_F/ D k) \coth(k L) $, where $k=1/\sqrt{D \tau_{sf}}$  .

The effects of the Luttinger liquid physics  are best observed by
estimating the renormalization of the Gilbert damping $\alpha$, and focusing
on its temperature dependence.
The spin pumping causes a negligible renormalization to $\gamma$, as
$\gamma \hbar/M_s \ll 1$. The corrections to the Gilbert damping  $\alpha$ 
are more significant, as $\alpha_0$ itself
is a small quantity \cite{Brataas1,Brataas2} (of the order of $10^{-2}$ in some
materials). In order to extract the temperature dependence of this correction, for illustrative 
purposes, we focus on the particular limit $\xi {\cal T} \ll 1$, other limits can be explored
as well. In this case the Gilbert damping
shows a power law dependence in temperature :
\be
\alpha=\alpha_0+\frac{\gamma \hbar}{M_s} 
\Big(\frac{T}{\epsilon_0}\Big)^{\delta} C,
\label{alpha}
\ee
where the constant $C=[t_1 +t_2 \xi \theta/4 \pi]/(1+\xi^2 
\theta^2/16 \pi^2)$, and $t_1$ and $t_2$ are temperature 
independent coefficients defined in relation to 
$A_1/\hbar=t_1 (T/\epsilon_0)^{\delta}$, and $A_2/\hbar=t_2 (T/\epsilon_0)^{\delta}$.
The renormalization of the Gilbert damping is significant  if $\alpha_0$ is smaller than
the correction term in Eq. (\ref{alpha}). This can be achieved experimentally
for some clean materials in which $\alpha_0$ is small, and 
the experimental signature of this effect
would be a temperature power law dependence of the measured Gilbert damping
coefficient.

%We can try to estimate it by analyzing a few cases. If $L^2\ll 
%v_F^2 \tau_{sf} \tau_{el}$, 
%which may not be a reasonable assumptions for nanotubes which are very clean,
%$\xi\approx v_F \tau_{sf}/L \gg 1$. For the opposite case which is more
%appropriate for nanotubes $\xi =\sqrt{\tau_{sf}/\tau_{el}} \gg 1$. 
%Thus, $\xi$ is large, but it is hard to estimate it exactly - hard to extract 
%more information.

\section{Ferromagnet -- Luttinger liquid -- metal junction}

We now consider a setup consisting of a ferromagnet with time dependent 
magnetization in contact with a Luttinger liquid of length $L$. The Luttinger
liquid is connected at the other end to a normal metal. We assume that
the length of the wire is shorter than the mean free path for spin flip 
so that the transport in the wire is ballistic. 

\begin{figure}[h]
\begin{center}
\includegraphics[width=2.5in]{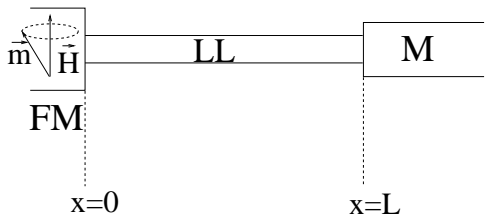}
\end{center}
\vspace{0.15in}
\caption{Ferromagnet-Luttinger liquid-metal}
\label{fig2}
\end{figure}

As before we can write down equations for the current
flowing through the wire. The current flowing from the FM to the LL is described
by the same equations as in the previous section, while the current
at the LL-metal junction is described by Ohm's law. We assume that the
transport in the wire is ballistic so that the spin chemical potential
$\vec{\mu}_s$ and the spin current $\vec{I}$ are constant throughout the wire. In the 
metal we assume that the spin dynamics is dominated by diffusion with $D_m$ being
the diffusion coefficient, and $\tau_{sf}^m$ being the spin flip time.
The details of the derivation are given in Appendix C.
We find 

\ba
\gamma'&=&\frac{\gamma}{1+\frac{\gamma}{M_s}(C_1 A_2-C_2 A_1)},
\nonumber \\ 
\alpha&=&\frac{\alpha_0+\frac{\gamma}{M_s}(C_1 A_1+ A_2  C_2)}{1+
\frac{\gamma}{M_s}(C_1 A_2-C_2 A_1)}.
\ea
Here $C_1=a_1/(a_1^2+a_2^2)$, $C_2=a_2/(a_1^2+a_2^2)$, 
\ba
a_1&=&\Big[1+\frac{{\cal T}_1}{{\cal T}_2}(1+{\cal T}_2 \xi_M)\Big], 
\\\nonumber
a_2&=&\frac{\theta}{4 \pi} \Big(\frac{1+{\cal T}_2 \xi_M}
{{\cal T}_2}\Big),
\ea
and $\xi_M=
 \sqrt{D_m \tau_{sf}^m}\coth(L/\sqrt{D_m \tau_{sf}^m}) /({\cal N}_m \hbar D_m
S)$.
Also ${\cal N}_m$ is the density of states in the metal, $D_m$
is the diffusion coefficient in the metal, and $S$ is the cross-section of the metal.
Again we find that the corrections to $\gamma$
are  negligible. We can further assume that the metal is a good spin sink, $\tau_{sf}^m$
is very small, and the term
proportional to $\xi_M$ can be dropped. Consequently we can write

\be
\alpha=\alpha_0+\frac{\gamma \hbar}{M_s} 
\Big(\frac{T}{\epsilon_0}\Big)^{\delta} C_m,
\ee
where
\be
C_m=\frac{t_1 (1+{\cal T}_1/{\cal T}_2)+t_2 (\theta/4 \pi {\cal T}_2)}{
(1+{\cal T}_1/{\cal T}_2)^2+(\theta/4  \pi  {\cal T}_2)^2}.
\ee
For illustrative purposes we consider two limits.
First, if $\theta=0$, $D=t_1/(1+{\cal T}_1/{\cal T}_2)$ and is temperature independent. Second,
if $\theta \gg 
{\cal T}_2$,
\be
\alpha=\alpha_0+\frac{\gamma \hbar}{M_s} 
\Big(\frac{T}{\epsilon_0}\Big)^{2 \delta} \frac{t_2 4 \pi {\cal T}_2^0}
{\theta}
\ee
is renormalized by a term which shows a temperature power law dependence 
with an exponent twice as large than in the previous case.
Here ${\cal T}_2^0$ is temperature independent
and is defined by  ${\cal T}_2= {\cal T}_2^0 (T/\epsilon_0)^{\delta}$.

\section{Conclusions}

We analyzed the injection of spin from a ferromagnet with
time dependent magnetization into a Luttinger liquid. We found that 
 the spin current injected into the LL has a similar form to the current
 that would be injected 
into a normal metal. However, the effect of the interactions in the LL is to 
suppress the spin pumping as a power law in temperature.
We  computed the total current flowing from a FM into a LL in
a few cases: a FM-LL-FM junction, a FM connected to a finite size
LL, and a FM-LL-metal junction. As a result of the LL physics, we found
that in these setups the
renormalization of the Gilbert damping is also suppressed from the 
normal metal case. It would be interesting to see if future experiments can
identify junctions in which this power law suppression is observed. Also 
it would be interesting to test for spin currents injected through this method
and transported over longer distances in good LL's (e.g. carbon nanotubes).

\section{Aknowledgements}
This work has been supported  by the NSF through grant DMR-9985255, and by the
Sloan and Packard Foundations. C.B. was also supported by the Broida-Hirschfeller
Foundation.

\begin{appendix}

\section{FM-LL-FM junction}

If we consider the exchange coupling to be negligible, Eq.(\ref{is}) describing
the spin current simplifies to

\be
\vec{I}=\frac{\vec{I}_0^1-\vec{I}_0^2}{2}.
\ee

We substitute this result for the spin current,
into the corresponding coupled LLG Eqs.(\ref{llg}) for the ferromagnets' magnetization
vectors. We assume that in each FM the magnetization is 
$\vec{m}_i=\hat{H}_i+\delta \vec{m}_i$, and $\delta \vec{m}_i$ are small, so that 
we  can write linearized equations in 
$\delta \vec{m}_i$. We consider the case two magnetic fields parallel to 
each other in the two ferrromagnets, $\vec{H}_1 || \vec{H}_2 || {\hat z}$. 
We note that up to quadratic 
corrections, the coupled equations preserve the magnitudes of $\vec{m}_i$,
i.e. $\vec{m}_i \cdot d \vec{m}_i/dt=0 $.

Thus the two coupled equations, 
\begin{widetext}
\ba
&&
\Big(1+\frac{\gamma A_2}{2 M_s^1}\Big)
\frac{d \vec{m}_1}{dt}=-{\gamma} \vec{m}_1\times \vec{H}^1+
\Big(\alpha_0+\frac{\gamma A_1}{2 M_s^1}\Big) \vec{m}_1  \times
\frac{d \vec{m}_1}{dt}+\frac{\gamma A_2}{2 M_s^1}\frac{d \vec{m}_2}{dt}-
\frac{\gamma A_1}{2 M_s^1}\vec{m}_2  \times
\frac{d \vec{m}_2}{dt}
\nonumber \\&&
\Big(1+\frac{\gamma A_2}{2 M_s^2}\Big)
\frac{d \vec{m}_2}{dt}=-{\gamma} \vec{m}_2\times \vec{H}^2+
\Big(\alpha_0+\frac{\gamma A_1}{2 M_s^2}\Big) \vec{m}_2  \times
\frac{d \vec{m}_2}{dt}+\frac{\gamma A_2}{2 M_s^2}\frac{d \vec{m}_1}{dt}-
\frac{\gamma A_1}{2 M_s^2}\vec{m}_1  \times
\frac{d \vec{m}_1}{dt}
\ea
\end{widetext}

can be linearized to yield

\begin{widetext}
\ba
&&(1+\alpha_2)
\frac{d \delta m_x^1}{d t}=- \gamma\delta m_y^1 H_1 -(\alpha_0+\alpha_1)
\frac{d\delta m_y^1}{d t}+\alpha_1\frac{d\delta m_y^2}{d t}+\alpha_2
\frac{d\delta m_x^2}{d t},
\nonumber \\&&
(1+\alpha_2)
\frac{d\delta m_y^1}{d t}=\gamma \delta m_x^1 H_1 +(\alpha_0+\alpha_1)
\frac{d\delta m_x^1}{d t}-\alpha_1\frac{d\delta m_x^2}{d t}+\alpha_2
\frac{d \delta m_y^2}{d t},
\nonumber \\&&
(1+\alpha_2)
\frac{d \delta m_x^2}{d t}=\gamma\delta m_y^2 H_2 -(\alpha_0+\alpha_1)
\frac{d \delta m_y^2}{d t}+\alpha_1\frac{d\delta  m_y^1}{d t}+\alpha_2
\frac{d \delta m_x^1}{d t},
\nonumber \\&&
(1+\alpha_2)
\frac{d\delta  m_y^2}{d t}=-\gamma\delta m_x^2 H_2 +(\alpha_0+\alpha_1)
\frac{d \delta m_x^2}{d t}-\alpha_1\frac{d \delta m_x^1}{d t}+\alpha_2
\frac{d\delta  m_y^1}{d t},
\ea
\end{widetext}
where $\alpha_{1/2}=\gamma A_{1/2}/ 2 M_s$ and we assumed that
the two ferromagnets are identical, so that $M_s^1=M_s^2=M_s$.
We can solve the above linear system of equations in the particular
case $\vec{H}_1=\vec{H}_2=H \hat{z}$. Assuming that the initial configuration 
is $\delta m_x^1=m_1$, $\delta m_x^2=m_2$, and $\delta m_y^{½}=0$, 
with $m_1\ll1$
and $m_2 \ll 1$, we obtain
\begin{widetext}
\ba
m_x^{1/2}&=&\frac{m_1+m_2}{2}\cos\frac{\gamma H t}{1+\alpha_0^2}
\exp\Big(-\frac{\alpha_0 \gamma
H t}{1+\alpha_0^2}\Big)
\nonumber \\&&
\pm \frac{m_1-m_2}{2}\cos\frac{(1+2 \alpha_2)\gamma H t}{(\alpha_0+
2\alpha_1)^2+(1+2 \alpha_2)^2}
\exp\Big[-\frac{(\alpha_0 +2 \alpha_1)\gamma H t}{(\alpha_0+
2\alpha_1)^2+(1+2 \alpha_2)^2}\Big],
\nonumber \\
m_y^{1/2}&=&\frac{m_1+m_2}{2}\sin\frac{\gamma H t}{1+\alpha_0^2}
\exp\Big(-\frac{\alpha_0 \gamma
H t}{1+\alpha_0^2}\Big)
\nonumber \\&&
\pm \frac{m_1-m_2}{2}\sin\frac{(1+2 \alpha_2)\gamma H t}{(\alpha_0+
2\alpha_1)^2+(1+2 \alpha_2)^2}
\exp\Big[-\frac{(\alpha_0 +2 \alpha_1)\gamma H t}{(\alpha_0+
2\alpha_1)^2+(1+2 \alpha_2)^2}\Big],
\ea
\end{widetext}

\section{FM-LL junction}

As described in Section II, the injected current is
\be
\vec{I_0}=-A_1\vec{m} \times \frac{d \vec{m}}{dt}+A_2 
\frac{d \vec{m}}{dt}.
\label{ia}
\ee
If we denote the chemical potential of the spin accumulated in the LL
 by $\vec{\mu}_s$,   
the backscattered current due to the accumulation of spin (Eq. (\ref{ib}) can be written as
\be
\vec{I}_b={\cal T}\vec{\mu}_s -\frac{\theta}{4 \pi} \vec{\mu}_s \times \vec{m},
\label{iba}
\ee
in the limit
$\mu_s \ll k_B T$.
The total current is given by $\vec{I}=\vec{I_0}-\vec{I}_b$, where $\vec{I}_0$ 
and $\vec{I}_b$
are given respectively by the equations (\ref{ia}) and (\ref{iba}).

 Along the lines 
of  Ref. ~\onlinecite{Brataas1} the diffusion 
equation for the spin in the LL is
\be
i \omega \vec{\mu}_s= D \partial_x^2 \vec{\mu}_s-\tau_{sf}^{-1} \vec{\mu}_s,
\ee
with the boundary conditions
$\partial_x \vec{\mu}_s(x)=- (2 \pi v_F/D)  \vec{I}$
at $x=0$,
and of vanishing spin current, $\partial_x 
\vec{\mu}_s(x)=0$, at $x=L$.

Correspondingly, the accumulated spin chemical potential in the LL at the 
junction with the FM is
$\vec{\mu}_s=\xi \vec{I}$, where $\xi=
(2 \pi v_F/D k) \coth(k L) $, $D$ is the diffusion coefficient in the wire,
and  $k=\sqrt{1+i \omega \tau_{sf}}/\sqrt{D 
\tau_{sf}}$.
Similar to Ref.~\onlinecite{Brataas1}, Since the frequency of precession of the
ferromagnet is much smaller than the inverse spin flip time, we can take
$k\approx 1/\sqrt{D \tau_{sf}}$, like in Ref.~\onlinecite{Brataas1}.

Consequently we obtain an equation for $\vec{I}$

\be
(1+\xi {\cal T}) \vec{I} -  \frac{\xi \theta}{4 \pi}
\vec{I} \times \vec{m}=  A_1 \vec{m} \times \frac{d \vec{m}}{d t}
- A_2 \frac{d \vec{m}}{d t}= \vec{I}_0,
\ee
whose solution is

\ba
\vec{I}&=&B_1  \vec{I}_0+B_2 (\vec{I}_0 \times \vec{m})
\nonumber \\&=&(B_1 A_2-B_2 A_1)
\frac{d \vec{m}}{d t}-(B_1 A_1+B_2 A_2) \vec{m} \times \frac{d \vec{m}}{d t},
\nonumber \\
\ea
where $B_1=
(1+\xi {\cal T}) /[(1+\xi {\cal T})^2+(\xi \theta)^2/(16 \pi^2)]$,
and $B_2=(\xi \theta)/\{4 \pi [(1+\xi {\cal T})^2+(\xi \theta)^2/(16 \pi^2)]\}$.
Here we made the assumptions that the length of the magnetization vector
$\vec{m}$ is not changing with time and for simplicity we took it to 
be $|\vec{m}|=1$; consequently $\vec{m} \cdot  (d \vec{m}/dt)=0$. 
The spin current $\vec{I}$ is then incorporated in the LLG 
equation
\be
\frac{d\vec{m}}{dt}=
-\gamma \vec{m}\times{\vec{H}}+\alpha_0 \vec{m}\times\frac{d \vec{m}}
{dt}
-\frac{\gamma}{M_s} \vec{I},
\label{gl}
\ee
which can be rewritten as

\be
\frac{d\vec{m}}{dt}=
-\gamma' \vec{m}\times{\vec{H}}+
\alpha \vec{m}\times\frac{d \vec{m}}{dt}.
\label{gl1}
\ee

Here $\gamma'=\gamma/[1+(\gamma/M_s)(B_1 A_2-B_2 A_1)]$, 
 and $\alpha=[\alpha_0+(\gamma/M_s)(B_1 A_1+ A_2  B_2)]/
[1+(\gamma/M_s)(B_1 A_2-B_2 A_1)]$.

\section{FM-LL-metal junction}

The current flowing from the ferromagnet to the LL is 
\be
\vec{I}=\vec{I}_0+
\frac{\theta}{4 \pi} \vec{\mu}_{s1} \times \vec{m}-{\cal T}_1\vec{\mu}_{s1}. 
\ee
The current flowing from the LL into the metal is
\be
\vec{I}={\cal T}_2(\vec{\mu}_{s1}-\vec{\mu}_{s2}),
\ee
where $\vec{\mu}_{s1}$ is the spin chemical potential in the wire, which
in case of ballistic transport is uniform along the LL, and $\vec{\mu}_{s2}$
is the spin chemical potential in the metal at point $B$. The two 
``spin conductances'' ${\cal T}_1$ and ${\cal T}_2$ are defined similarly
to the previous appendix, with the index $1$ corresponding to 
the FM-LL interface, and the index $2$ to the LL-metal interface.
The spin current  and the spin chemical potential
 $\vec{\mu}_{s2}$  in the metal can  be related by \cite{Brataas1}
 $\vec{\mu}_{s2}=\xi_M \vec{I}$. Here $\xi_M=\coth(k_m L)/({\cal N}_m \hbar
D_m S k_m) $, ${\cal N}_m$ is the density of states in the metal, $D_m$ is the diffusion 
coefficient in the metal, $S$ is the
cross-section of the metal, and $k_m=\sqrt{1+i \omega \tau_{sf}^m}/
\sqrt{D_m \tau_{sf}^m}\approx 1/\sqrt{D_m \tau_{sf}^m}$.
The corresponding equation for the spin current flowing through the junction
is 

\be
a_1 \vec{I} -a_2
\vec{I} \times \vec{m}=  \vec{I}_0,
\ee
with 
\be
a_1=\Big[1+\frac{{\cal T}_1}{{\cal T}_2}(1+{\cal T}_2 \xi_M)\Big],
\ee
and
\be
a_2=\frac{\theta}{4 \pi} \Big(\frac{1+{\cal T}_2 \xi_M}
{{\cal T}_2}\Big).
\ee
Solving for $\vec{I}$ we get
\ba
\vec{I}&=&C_1  \vec{I}_0+C_2 (\vec{I}_0 \times \vec{m})
\nonumber 
\\&=&(C_1 A_2-C_2 A_1)
\frac{d \vec{m}}{d t}-(A_1 C_1+C_2 A_2) \vec{m} \times \frac{d \vec{m}}{d t},
\nonumber \\
\ea
where $C_1=a_1/(a_1^2+a_2^2)$ and $C_2=a_2/(a_1^2+a_2^2)$.
We thus find in this case 
$\gamma'=\gamma/[1+(\gamma/M_s)(C_1 A_2-C_2 A_1)]$, 
 and $\alpha=[\alpha_0+(\gamma/M_s)(C_1 A_1+ A_2  C_2)]/
[1+(\gamma/M_s)(C_1 A_2-C_2 A_1)]$ .
\end{appendix}

%\end{multicols}

\begin{thebibliography}{99}
%\bibitem{Brataas1} Y. Tserkovnyak, A. Brataas and G. E. W. Bauer,
%Phys. Rev. B 66, 224403 (2002).
%\bibitem{Brataas2}  Y. Tserkovnyak, A. Brataas and G. E. W. Bauer, Phys.
%Rev. Lett. {\bf 88}, 117601 (2002). 
\bibitem{inj} G. Schmidt, D. Ferrand, D. W. Molenkamp, A. T. Filip and
B. J. van Wees, Phys. Rev. B {\bf 62}, R4790 (2000).
\bibitem{optical} J. M. Kikkawa and D. D. Awschalom, Nature
{\bf 397}, 139 (1999).
\bibitem{pump} P. W. Brouwer, Phys. Rev. B {\bf 58} R10135 (1998).
\bibitem{Brataas1} Y. Tserkovnyak, A. Brataas and G. E. W. Bauer,
Phys. Rev. B 66, 224403 (2002).
\bibitem{Brataas2}  Y. Tserkovnyak, A. Brataas and G. E. W. Bauer, Phys.
Rev. Lett. {\bf 88}, 117601 (2002). 
\bibitem{Brataas3} A. Brataas, Y. Tserkovnyak, G. E. W. Bauer and
B. I. Halperin, Phys. Rev. B 66, 060404 (2002)
\bibitem{Gilbert} T. L. Gilbert, Phys. Rev. {\bf 100}, 1243 (1955);
L. D. Landau, E. M. Lifshitz and L. P. Pitaevski, 
{\it Statistical Physics, part 2} (Pergamon, Oxford, 1980), 3rd ed.
\bibitem{KF} C. L. Kane and M. P. A. Fisher, Phys. Rev. Lett. {\bf 68}, 1221 (1994);
Phys. Rev. B, {\bf 46}, 15233 (1992).
\bibitem{spin1} L. Balents and R. Egger, Phys. Rev. B 64, 035310 (2001).
\bibitem{spin2} L. Balents, 1999 Moriond Les Arcs Conference Proceedings,
cond-mat/9906032.
\bibitem{spin3} L. Balents and R. Egger, Phys. Rev. Lett. 85, 3464-3467 (2000).
\end{thebibliography}
\end{document}